# 5G Fronthaul – Latency and Jitter Studies of CPRI over Ethernet

Divya Chitimalla[1], Koteswararao Kondepu[2, 3], Luca Valcarenghi[2], Massimo Tornatore[1], and Biswanath Mukherjee[1].

*[1]University of California Davis, USA; [2]Scuola Superiore Sant' Anna, Italy*
*[3]High Performance Networks Group, University of Bristol, UK*
{dchitimalla, mtornatore, bmukherjee}@ucdavis.edu, {k.kondepu, luca.valcarenghi}@sssup.it

*Abstract—* **Common Public Radio Interface (CPRI) is a successful industry cooperation defining the publicly-available specification for the key internal interface of radio base stations between the Radio Equipment Control (REC) and the Radio Equipment (RE) in the fronthaul of mobile networks. However, CPRI is expensive to deploy, consumes large bandwidth, and currently is statically configured. On the other hand, Ethernet-based mobile fronthaul will be cost-efficient and more-easily reconfigurable. Encapsulating CPRI over Ethernet (CoE) is an attractive solution, but stringent CPRI requirements such as delay and jitter are major challenges that need to be met to make CoE a reality. This study investigates whether CoE can meet delay and jitter requirements by performing FPGA-based Verilog experiments and simulations. Verilog experiment shows that CoE encapsulation with fixed Ethernet frame size requires about tens of microseconds. Numerical experiments show that the proposed scheduling policy of CoE flows on Ethernet can reduce jitter when redundant Ethernet capacity is provided. The reduction in jitter can be as large as one microsecond hence making Ethernet-based mobile fronthaul a credible technology.**

## I. INTRODUCTION

Extensive adoption of smart phones and smart devices has enormously increased bandwidth consumption in cellular networks [1], thus calling for effective ways to improve cellular capacity. For example, 5G bandwidth consumption is expected to be 1000x of 4G [1] [2], which calls for novel Radio Access Network (RAN) architectures that can support much higher bandwidths in a cost-effective manner. A popular approach is to split the functionalities of 4G evolved NodeB (eNB) into a radio equipment (RE), consisting of antenna and basic radio frequency (RF) functionality, and radio equipment controller (REC), which processes the signals from the physical layer and above. This solution was originally called Centralized RAN (C-RAN) as multiple RECs could be consolidated in a single centralized location, and single REC can be shared among many REs, depending on traffic load. C-RAN can significantly increase the cellular coverage density by deploying many REs which are lightweight compared to full-fledged macro base stations, and thereby reduce network cost by using fewer RECs. Recent proposals push the REC function into the "cloud" (where the REC is "virtualized"), thereby moving from Centralized-RAN to Cloud-RAN and Virtualized-RAN (V-RAN) [3].

Several ongoing projects, such as the Institute of Electrical and Electronics Engineers (IEEE) Standards Association 1914.1 working group [4], are striving to define an interface (electrical, optical, or wireless) between REC and RE. The interface requirements depend on the functional split [5] which, as proposed by 3GPP, is the set of functionalities that exist in the RE and REC. The split can occur at several protocol layers, thus resulting in different bandwidth and delay requirements of the mobile fronthaul. Our study considers the split at the physical layer of eNB (i.e., Option 8 in TR 38.801), which includes the entire layer 1 and above functions in the REC, whereas RE is a lightweight antenna having only RF functionality. In this option in-phase quadrature (IQ) samples of the baseband signal must be transported between RE and REC. Common Public Radio Interface (CPRI) is a well-known radio interface developed by several leading telecom vendors to transport sampled RF data between RE to REC. CPRI is a constant-bit-rate (CBR) interface with line rate options ranging from 614.4 Mbps (option 1) up to 24.33 Gbps (option 10) [6]. CPRI is a product of industry cooperation which is of a closed nature, while other interfaces of more open nature exist (e.g., Open Base Station Architecture Initiative (OBSAI) and Open Radio Equipment Interface (ORI)) [7] [8].

CPRI is manufactured in low volumes, thus making it expensive. It is also extremely difficult to design switching equipment for CPRI. Although CPRI mentions that it supports several topologies such as tree, ring, and chain [6], there is no mention on how these topologies can be controlled. CPRI has stringent delay and jitter requirements, which can be satisfied only with high-speed fronthaul solutions (e.g., optical links) as in [9]. All these issues make it imperative to design a cost-efficient and reconfigurable mobile fronthaul that supports emerging network paradigms.

Encapsulating CPRI over Ethernet (CoE) is a cost-efficient solution which can leverage existing Ethernet interfaces, and switching equipment, for mobile fronthaul. Ethernet has many advantages such as easy upgradability to higher data rates, wide-scale availability, low-cost equipment, and ease of scalability. Moreover, Ethernet switches can be used to configure a fronthaul into any network topology, even on a large network scale. Another advantage of utilizing CoE is that current high-speed optical networks can also be utilized for mobile fronthaul. In particular, 10 Gigabit (10G) Ethernet is fast enough to carry high-data-rate sampled IQ signals from REC to RE (e.g., a 20-MHz single-antenna I/O sampled radio signal can be handled by 10G Ethernet interface). Transport options such as dedicated fiber, optical transport network (OTN), and passive optical network (PON) [10] can support the fronthaul by deploying fibers and other optical components (e.g., switches, Optical Line Terminals (OLTs) for eNB-to-eNB communication). However, whether Ethernet can support stringent CPRI requirements in terms of delay and jitter is under scrutiny as the Ethernet mobile fronthaul needs to support delay within 100 µs and jitter within 65 ns [6], among other strict requirements of the time-sensitive IQ data that is being transmitted.

An ongoing effort by industry [11] [12] and academia [13-15] is investigating Ethernet fronthaul solution. The IEEE Standard Association (SA) 1914 working group is effective since 2015 to standardize Radio over Ethernet (RoE) [4]. In particular, IEEE 1914.3 task force is investigating ways of transferring IQ user-plane data, vendor-specific data, and control and management (C&M) information channels [6] over an Ethernet-based packet-switched network. This standard focuses on encapsulating data into the Ethernet frame payload field with an additional RoE header for timing and synchronization purposes. Two types of encapsulation are defined in RoE: structure-aware and structure-agnostic. Structure-aware encapsulation uses knowledge of the encapsulated and digitized radio transport format content, whereas structure-agnostic encapsulation is a container that encapsulates bits into Ethernet frames irrespective of the encapsulated protocol. The applicability of Ethernet to mobile fronthaul has been discussed in [11] by exploiting the buffers to reduce the jitter of Ethernet packets. However there is no experimental or simulative study quantifying the jitter in the proposed Ethernet fronthaul implementation. Several TSN Ethernet techniques (e.g., 802.1Qbu Frame Preemption and 802.1Qbv with guard band) for carrying fronthaul data has been compared in [12], however, there is no detailed study on under which conditions Ethernet with scheduled traffic can achieve less than 65 ns. Moreover, to minimize jitter in Ethernet fronthaul, scheduling Ethernet frames with fixed timeslots to a specific flow has been proposed in [13]. A functional split between the REC and RE has been proposed in [14] that permits baseband signal transport instead of the transport of sampled radio streams, to enable lower-rate fronthaul. Such a fronthaul can also make use of Ethernet switches, and networking statistical multiplexing gains, as it transports relatively bursty data instead of continuous radio waveforms. Furthermore, [15] provides experimental realization of dynamically reconfigurable CPRI over Ethernet, and also provides delay analysis of dynamically reconfigurable Ethernet fronthaul.

There are investigations within IEEE 802.1CM whether IEEE 802.1Qbu [16] and IEEE 802.1Qbv [17] using preemption and scheduling could be utilized to guarantee latency and jitter requirements for Ethernet fronthaul. IEEE 802.1Qbu is utilizing frame preemption policies where IEEE 802.3br provides the mechanism to implement preemption at the media access control (MAC) and below layers. IEEE 802.1 Qbv is working on scheduled traffic with edge buffer which absorbs variation in packet delay with the added delay cost. The works in [18] [19] provide enhancements to IEEE 802.1Qbu and IEEE 802.1Qbv standards. These studies have shown that (i) by using 802.1Qbu pre-emption in Ethernet cannot meet jitter requirements of 65 ns and (ii) 802.1Qbv using Ethernet scheduling can remove jitter in some cases depending on the input flows, but not always. 802.1Qbv utilizes guard bands to absorb fluctuations in the schedule of Ethernet packets. The size of the guard band determines the performance of 802.1Qbv Ethernet, where small guard band size increases packet collisions and large guard band size decreases the effective throughput of Ethernet. IEEE 802.1Qbv and IEEE 802.1Qch address the synchronization problems such as latency and jitter in networks where time-sensitive data shares capacity along with non-time-sensitive data. In particular, IEEE 802.1Qch describes the methods that can be adopted to schedule flows at strict time intervals using on-off gates for scheduled Ethernet. IEEE 802.1Qbv enhances the methods suggested in 802.1Qch to include VLAN tags to prioritize time-sensitive traffic such that delay/jitter get reduced. Our work assumes that the fronthaul network is capable of implementing the methods as described by Qch and Qbv. However these standards do not explicitly describe any algorithm to minimize jitter in Ethernet fronthaul. In this work, we provide a scheduling algorithm for CoE data such that jitter remains within 65 ns for the given CoE data rates, which is not specified in Qbv/Qch. We also estimate the Ethernet capacity required to achieve tolerable jitter (65 ns) for a given set of CoE flows in the Ethernet fronthaul.

Our study provides a quantitative performance evaluation of CoE in terms of delay and jitter. An FPGA pre-synthesis evaluation is performed to verify the logical functionality of CoE design and encapsulation overhead. Moreover, we exploit advances in time-sensitive networking (TSN) such as scheduling Ethernet (IEEE 802.1Qbv) to devise an exhaustive-search algorithm that returns jitter-reduced frame scheduling.

The rest of the study is organized as follows. In Section II, we discuss the CoE-based mobile fronthaul architecture. In Section III, we give the mapping between CPRI and Ethernet frames, where we also evaluate its important parameters such as encapsulation delay, Ethernet overhead and distance supported by Ethernet fronthaul. Section IV discusses jitter-minimization techniques for CoE. We propose algorithms that can be programmed in the Ethernet switch that reduce jitter in Ethernet fronthaul. In Section V, we perform Verilog experiments and simulations to evaluate the delay and jitter of CoE-based fronthaul. Section VI concludes the study.

## II. CoE-Based C-RAN Architecture

### A. Frame Structure

CPRI sends sampled IQ data in a frame format as shown in Fig. 1. It uses fixed-bandwidth connections between REC and RE with different line rates (options 1 to 10) [6]. CPRI supports 8B/10B and 64B/66B encoding options; without loss of generality, our study considers 8B/10B encoding. While CPRI supports topologies such as tree, ring, and chain, each link between RE and REC is a fixed-bandwidth time-division-multiplexed (TDM) connection. A single basic frame duration is 260 ns (1/3.84 MHz) which is compatible to a Universal Mobile Telecommunications System (UMTS) chip length. Each basic frame consists of 16 words, and the word length depends on the CPRI line rate [6]: 256 basic frames make a hyper frame, and 150 hyper frames make a radio frame.

The CPRI radio frame is 10 ms. CPRI line rate information is sent in Z.Y.W.X format between RE and REC, where Z is hyper frame number, Y is basic frame within a hyper frame, W is word number within a basic frame, and X is byte number within a word. CPRI provides auto-rate negotiation which allows a dynamic reconfiguration of the CPRI line rate based on the antenna, and hence user traffic characteristics [6].

### B. Network Architecture

The considered architecture for Ethernet-based C-RAN and/or V-RAN is shown in Fig. 2 where there are three links from RE to REC supporting CPRI flows packetized over Ethernet. The CoE flows from RE to REC pool are switched using Ethernet switch (SW), where a scheduling policy can be programmed to provide access control to avoid collision.

This architecture can support network sharing between multiple vendors and operators as envisioned for 5G network [20], which should integrate different wireless standards: 3G, 4G, LTE-advanced, and WiFi. Also, several physical media can be used based on the demand and availability of resources such as fiber, cable, DSL, mm-wave, and free-space optics. The proposed architecture can jointly provision resources of different media for fronthaul and backhaul (the connection between REC pool to core network), as Ethernet can be the underlying protocol for each of these platforms. This facilitates network sharing and common operation and maintenance (OAM) functions. If large capacity is requested optical infrastructure can be utilized to support the RAN [21]. Moreover, thanks to virtualization, multiple operators can share a common physical infrastructure [22].

## III. CPRI-over-Ethernet (CoE) Mapping

CoE encapsulation requires a mapping between CPRI and Ethernet frames. In this study, we describe a structure-agnostic mapping of CoE where CPRI flows are sequentially packetized onto an Ethernet frame without the knowledge of CPRI data. Several CoE flows can share a common Ethernet link. Table I shows the notations utilized to describe the mapping between CPRI and Ethernet frames. Figure 3 shows the encapsulation of CPRI flows in the payload of Ethernet frames, considering that the input is at CPRI line rate and output is at Ethernet link rate. The CPRI data is framed into Ethernet with additional MAC and RoE header: preamble (7 bytes), start of frame delimiter

(1 byte), source address (6 bytes), destination address (6 bytes), Ethernet type (2 bytes), RoE header (6 bytes), frame check sequence (4 bytes), and inter-packet gap (12 bytes). As in [4], 6 bytes of RoE header further contains different sub-fields such as: version, packet type, start of the frame, flow id, timestamp select field, timestamp, and optional extended RoE header space. Note that optional 802.1Q tag field is not considered in Ethernet overhead calculations. The CPRI data in an Ethernet frame is always a multiple of CPRI basic frame. The CoE mapping parameters such as encapsulation delay, hop delay, and Ethernet overhead are discussed below.

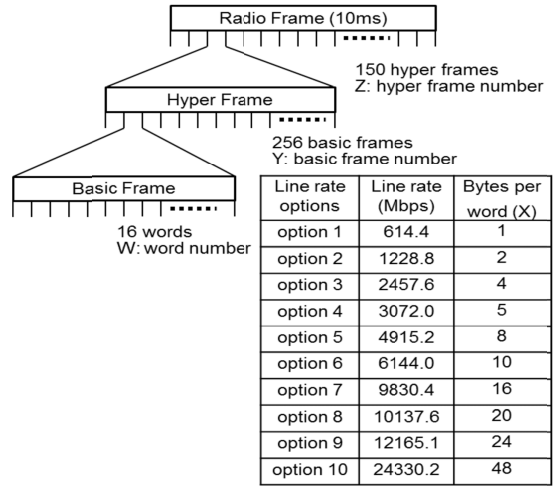

Figure 1: Frame structure of CPRI.

| Line rate options | Line rate (Mbps) | Bytes per word (X) |
|---|---|---|
| option 1 | 614.4 | 1 |
| option 2 | 1228.8 | 2 |
| option 3 | 2457.6 | 4 |
| option 4 | 3072.0 | 5 |
| option 5 | 4915.2 | 8 |
| option 6 | 6144.0 | 10 |
| option 7 | 9830.4 | 16 |
| option 8 | 10137.6 | 20 |
| option 9 | 12165.1 | 24 |
| option 10 | 24330.2 | 48 |

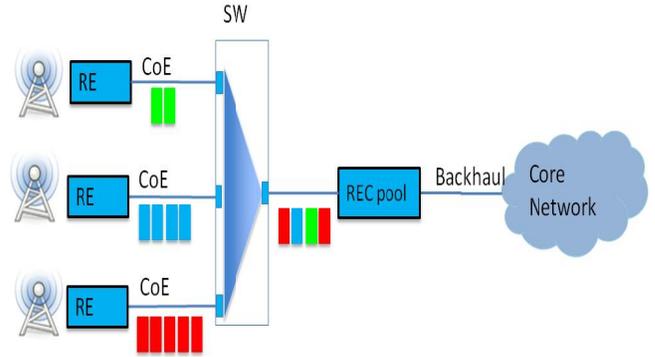

Figure 2: CPRI-over-Ethernet fronthaul architecture for C-RAN.

CoE encapsulation/de-capsulation is assumed to be performed at both RE and REC, and the minimum CPRI data to be encapsulated into Ethernet is one CPRI basic frame of duration $T_B \approx 260$ ns. Thus, Ethernet payload size $L_P$ is computed as:

$$L_P = N_B \cdot R_{CPRI} \cdot T_B \qquad (1)$$

The value of $N_B$ for different line rates is chosen such that the payload value remains close to 1250 or 1500 bytes. $L_P$ is made multiple of $N_B$ so the number of CPRI basic frames in an Ethernet frame remains an integer value, thus basic frame fragmentation is avoided.

Using $L_P$, the encapsulation delay $T_{encap}$ is defined as the time taken to frame the CPRI data at a specific line rate into Ethernet payload (i.e., time to receive the CPRI payload):

$$T_{encap} = \frac{L_P}{R_{CPRI}} = N_B \cdot T_B \qquad (2)$$

Total Ethernet header overhead ($T_{totHOH}$) is the additional delay to transmit Ethernet header ($L_{EH}$) bytes due to Ethernet encapsulation, and it depends on the total number of Ethernet frames ($N_E$) utilized to encapsulate the CPRI data, which depends on $R_{CPRI}$ and $L_P$:

$$T_{totHOH} = N_E . L_{EH}/R_E = N_E . T_{EOH} \qquad (3)$$

where $L_{EH}$ is set to a fixed value (i.e., 44 bytes), $R_E$ is the Ethernet line rate (e.g., line rate of 10G Ethernet: 10 Gbps), $T_{EOH}$ is the header overhead per Ethernet frame.

Hop (i.e., switch, router) delay ($T_{hop}$) is the delay introduced by the Ethernet switch to process the packet using a store-and-forward mechanism, when RE and REC are in multi-hop configuration [6]. Hop delay can be estimated based on the switch forwarding functionality which could be store-and-forward or cut-through mechanism. This paper considers a worst-case hop delay utilizing a store-and-forward switch. Cut-through switch reduces the hop delay (to 6.4ns compared to 1µs for store-and-forward) as only the first 8 bytes are needed to be processed before switch forwards the Ethernet packet to the respective output port:

$$T_{hop} = L_E/R_E \qquad (4)$$

where $L_E$ is length of Ethernet frame ($L_P + L_{EH}$), expressed in multiples of CPRI basic frame length $T_B$ (see Fig. 3).

TABLE I
NOTATIONS

| | |
|---|---|
| Length of Basic CPRI Frame[second] | $T_B$ |
| Length of Ethernet Frame[bit] | $L_E$ |
| Encapsulation Delay[second] | $T_{encap}$ |
| Ethernet Payload Size[bit] | $L_P$ |
| CPRI Line Rate[bit per second] | $R_{CPRI}$ |
| Header Overhead per Ethernet Frame[second] | $T_{EOH}$ |
| Total Ethernet Header Overhead[second] | $T_{totHOH}$ |
| Ethernet Header Size[bit] | $L_{EH}$ |
| Number of CPRI Basic Frames | $N_B$ |
| Number of Ethernet Frames in a Radio Frame | $N_E$ |
| Total CoE Overhead[second] | $T_{totEOH}$ |
| Ethernet Rate[bit per second] | $R_E$ |
| Hop Delay[second] | $T_{hop}$ |

From Eqs. (3) and (4), total CoE overhead ($T_{totEOH}$) caused by encapsulation of CPRI data on Ethernet is computed as:

$$T_{totEOH} = T_{totHOH} + T_{hop} \qquad (5)$$

Combining Eqs. (3)-(5), we get:

$$T_{totEOH} = N_E . L_{EH}/R_E + L_E/R_E$$

where Total Ethernet header overhead ($T_{totHOH}$) is the additional delay to transmit Ethernet header ($L_{EH}$) bytes due to Ethernet encapsulation, and it depends on the total number Ethernet frames ($N_E$) utilized to encapsulate the CPRI data. Hop (i.e., switch, router) delay ($T_{hop}$) is the delay introduced by the Ethernet switch to process the packet using a store-and-forward mechanism.

Table II shows the computed values of CoE parameters based on Eqs. (1)-(5) when 10G Ethernet is used to send CPRI line rates from option 1 (614.4 Mb/s) to option 6 (6144.0 Mb/s) for two Ethernet payloads sizes, $L_P$ of 1250 bytes and 1500 bytes. $T_{hop}$ values in Table II are for a single hop. They are critical to analyze the delay performance of CoE. LTE radio frame of 10 ms is divided into 10 sub-frames, each of 1 ms. A LTE eNB should complete eNB processing (uplink CPRI processing, uplink frame decoding, ACK/NACK creation, downlink frame creation, downlink CPRI processing) within 3 ms after receiving uplink data from User Equipment (UE) as the HARQ protocol needs an

ACK/NACK to be sent in 3 ms for every four LTE sub-frames. Hence, $T_{totEOH}$ for transmitting four sub-frames is also shown in Table II. Note that $T_{totEOH}$ is obtained by adding $T_{totHOH}$ for four sub-frames (4 ms) with single $T_{hop}$. From [23], the maximum allowed fiber round-trip time is 246 µs after removing RF processing time (40 µs), CPRI processing time (10 µs), REC processing time (2700 µs), and fronthaul equipment processing (4 µs) from 3 ms delay requirement. Thus, maximum distance supported (km) between REC and RE by CoE is given by:

$$Distance = (246µs - T_{totEOH})/10µs/km \qquad (6)$$

where 10 µs/km is round-trip fiber propagation delay as the speed of light in fiber is 200000km/s. Virtualized RECs can move across different REC pools (hotel of RECs that share cooling and housing resources to save energy) according to traffic/network requirements. This can lead to a situation where fronthaul data traverses different Ethernet switches, leading to a multi-hop scenario as explained in CPRI [6], where each hop corresponds to an Ethernet switch. Experiments conducted in the next section investigate the scheduling policies to reduce jitter in Ethernet fronthaul.

## IV. JITTER STUDY OF COE

Proper scheduling that minimizes jitter is crucial to achieve acceptable jitter performance on Ethernet fronthaul. An attractive solution to minimize jitter in Ethernet fronthaul, is scheduling Ethernet frames by assigning fixed timeslots to send packets of a specific CoE flow [13]. Figure 4(a) shows an example where three CoE flows (1, 2, 3) of rates 5000 Mbps, 2500 Mbps, and 1250 Mbps (each of $L_E$ = 1000 bytes) respectively are multiplexed on an Ethernet interface at 10 Gbps. *Scheduling length* is defined as shortest time interval where CoE packets are multiplexed whose pattern repeats periodically; in Fig. 4, scheduling length is denoted by $L_S$.

The difference in the inter-arrival time between packets is measured as the packet-to-packet jitter [24] [25]. The CoE input packets are isochronous meaning packets arrive the input of Ethernet switch at regular intervals. Inter-arrival jitter is usually taken as the absolute value of the deviation from its regular state. For evaluating jitter characteristics of fronthaul we take the worst case jitter value for all the CoE flows multiplexed as follows:

$$delay_{i,j} = arrival\ time_{i+1,j} - arrival\ time_{i,j}$$
$$Jitter_j = max_{\forall i}\ delay_{i,j} - min_{\forall i}\ delay_{i,j}$$
$$Jitter = max_{\forall j}\ Jitter_j \qquad (7)$$

where $delay_{i,j}$ denotes delay at the receiver of REC for packet number $i$ in flow $j$. From Eq. (7) worst case jitter for all flows (i.e., max of max) is taken as a quality metric of the schedule. For this example, jitter on flow 1 is the difference between highest inter-packet delay, i.e., 2.4 µs, and lowest inter-packet delay, i.e., 0.8 µs, which is 1.6 µs. However, a better scheduling can be done that completely removes jitter, as shown in Fig. 4(b), where the jitter is zero since there is no variance in inter-packet delay for packets of the same flow. In this section, we propose a scheduling policy to multiplex several CoE flows on Ethernet such that jitter of CoE remains within acceptable level (see Eq. (7)).

This scheduling policy can be programmed in the Ethernet switch shown in Fig. 2, where multiplexing occurs. CoE flows from several REs need to be scheduled at precise times to provide least delay variance, and hence tolerable jitter. Scheduling Ethernet requires strict (and periodic) time schedules (on/off slots) where each CPRI flow's packets can be transmitted. Schedule is formed using parameters such as queue schedule of nodes, transmission delays, packet lengths, and CoE rates.

A conflicting schedule is defined as one which schedules more than one packet of different or same flows at the same time. Finding a non-conflicting schedule of packets is proven to be NP-complete. Refs. [26] [27] prove that the problem of producing a non-conflicting schedule to multiplex multiple flows can be reduced to the classical graph-coloring problem, which is known to be NP-complete. There are several algorithms proposed in other network problems that strive to produce non-conflicting schedule of multiplexed packets [26] [27]; node-based scheduling and level-based scheduling are popular ones. But for fronthaul where topologies are not as complex as other networks, and jitter is much more stringent, a greedy approach that exhaustively searches the minimum jitter sequence can be a good choice. Below, we propose a greedy scheduling algorithm that minimizes jitter by proper scheduling, and then we compare it with a benchmark algorithm. We assume that all the flows in the proposed fronthaul network are CoE flows whose characteristics such as packet lengths and CPRI rates are well-determined. We also assume that the network is not oversubscribed and there is only one switch that is multiplexing multiple CPRI flows onto Ethernet output using a tree topology. If there are multiple switches aggregating flows in the network, a combined schedule needs to be formed using global information with the help of Software Defined Network (SDN) controller and pushed into each of the switches.

When multiple input ports get aggregated into an output port there is an internal serialization delay in the switch known as the M:1 delay. The objective of the proposed scheduling policies is to decrease the maximum jitter among all the flows as defined in the paper thus providing the frames with a fair amount of serialization delay.

### A. CoE Scheduling Policies

This section introduces the proposed *Comb-fitting (C-FIT)* algorithm that schedules flows in Ethernet to reduce jitter. *Basic-offset* algorithm provides an initial configuration to be used by C-FIT, and *First Available Timeslot (FAT)* serves as a benchmark algorithm. Table III shows several parameters which are utilized in the pseudo code of these algorithms. The basic-offset algorithm (Algorithm 1) schedules CoE packets such that jitter is temporarily zero (ideal case) without taking into consideration that the obtained solution can contain scheduling conflicts (i.e., multiple packets can be scheduled at the same time). CoE packets are offset by multiples of Ethernet timeslot sizes $T_{ETS}$ for each flow.

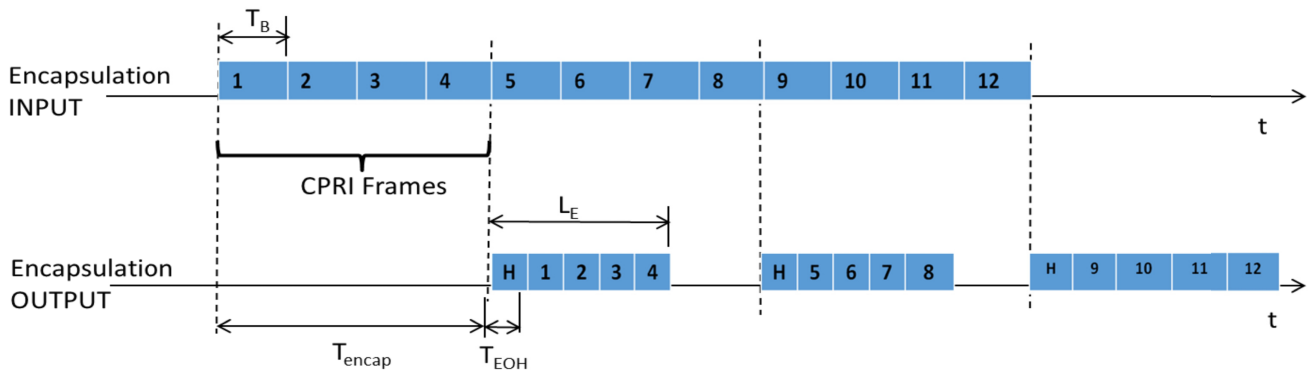

Figure 3: CPRI encapsulation over Ethernet.

TABLE II
CoE PARAMETERS

| Line rate [Mb/s] | Ethernet Packets per Radio Frame ($L_P$ =1250 bytes) | $T_{encap}$ [µs] | $T_{hop}$ [µs] | $T_{totHOH}$ (for radio frame) [µs] | $T_{totHOH}$ (for four sub-frames) [µs] | $T_{totEOH}$ *2 (Round Trip) [µs] | Distance supported [km] |
|---|---|---|---|---|---|---|---|
| 614.4 (option 1) | 615 | 16.27 | 1.00 | 21.65 | 8.66 | 18.32 | 22.77 |
| 1228.8 (option 2) | 1229 | 8.13 | 1.00 | 43.26 | 17.30 | 35.61 | 21.04 |
| 2457.6 (option 3) | 2458 | 4.06 | 1.00 | 86.52 | 34.61 | 70.22 | 17.58 |
| 3072.0 (option 4) | 3073 | 3.25 | 1.00 | 108.17 | 43.27 | 87.54 | 15.85 |
| 4915.2 (option 5) | 4916 | 2.03 | 1.00 | 173.04 | 69.22 | 139.43 | 10.66 |
| 6144.0 (option 6) | 6144 | 1.62 | 1.00 | 216.27 | 86.51 | 174.02 | 7.20 |
| Line rate [Mb/s] | Ethernet Packets per Radio Frame ( $L_P$ =1500 bytes) | $T_{encap}$ [µs] | $T_{hop}$ [µs] | $T_{totHOH}$ (for radio frame) [µs] | $T_{totHOH}$ (for four sub-frames) [µs] | $T_{totEOH}$ *2 (Round Trip) [µs] | Distance supported [km] |
| 614.4 (option 1) | 512 | 19.53 | 1.20 | 18.02 | 7.21 | 15.62 | 23.04 |
| 1228.8 (option 2) | 1024 | 9.76 | 1.20 | 36.04 | 14.42 | 30.04 | 21.60 |
| 2457.6 (option 3) | 2048 | 4.88 | 1.20 | 72.09 | 28.84 | 58.87 | 18.71 |
| 3072.0 (option 4) | 2560 | 3.90 | 1.20 | 90.11 | 36.04 | 73.29 | 17.27 |
| 4915.2 (option 5) | 4096 | 2.44 | 1.20 | 144.18 | 57.67 | 116.54 | 12.95 |
| 6144.0 (option 6) | 5120 | 1.95 | 1.20 | 180.22 | 72.09 | 145.38 | 10.06 |

*C-FIT* takes the schedule produced by *basic-offset algorithm* as input and resolves conflicts. All possible permutations of input flow orders are formed, since the order in which any two flows are combined using *matcombine* subroutine affects the final jitter. For example, if three flows are considered, then the possible flow orders are {1 > 2 > 3}, {1 > 3 > 2}, {2 > 1 > 3}, {2 > 3 > 1}, {3 > 1 > 2}, {3 > 2 > 1}. For each order, the flows are sequentially combined using *matcombine*. *matcombine* takes two flows' schedules as input and produces a non-conflicting schedule according to this procedure: the flow with higher number of packets is kept intact and the other flow is offset by a multiple of $L_{ETS}$ to produce a non-conflicting schedule (this is called sliding approach). If such a non-conflicting sequence is not achieved by using the sliding approach, conflicting packets are moved to the nearest timeslot that is unoccupied. This approach is followed for all possible flow orders and the schedule with lowest amount of jitter is selected as the final schedule.

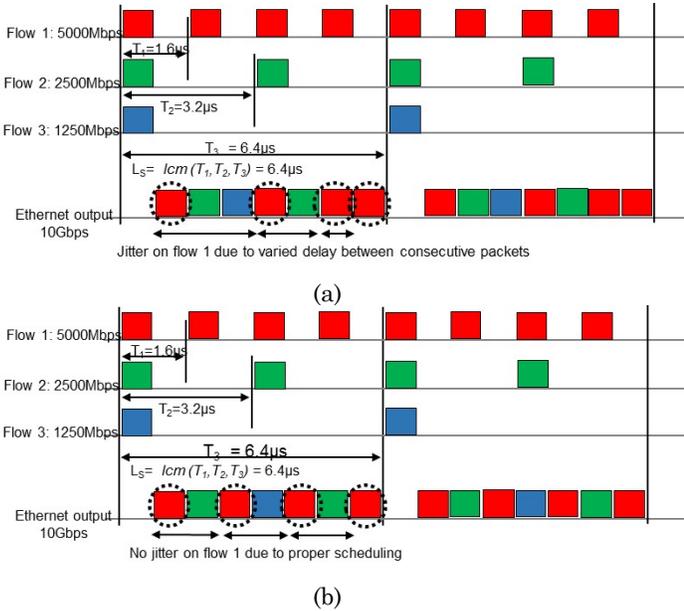

Figure 4: (a) An example that shows jitter on flow 1; (b) An example that shows how proper scheduling can eliminate jitter.

The proposed *C-FIT algorithm* is compared with *first-available-timeslot (FAT) algorithm*, namely benchmark *FAT*, which resolves the conflicts produced with *basic-offset algorithm* by moving the conflicting packets to the first available timeslot that can accommodate the packet without using sliding approach and flow ordering. Algorithm 3 shows the pseudo code for *benchmark FAT*.

TABLE III
PARAMETERS FOR CoE PACKET SCHEDULING

| Input CoE rate for flow $i$ [bit per second] | $R_{CoE}^i$ |
|---|---|
| Ethernet rate [bit per second] | $R_E$ |
| Transmission time of flow $i$ packet on Ethernet link [second] | $T_P^i$ |
| Ethernet timeslot size [second] | $T_{ETS}$ |
| # of slots in a schedule length for flow $i$ | $L_{SF}^i$ |
| Schedule of CoE flow $i$ on Ethernet link [timevector] | $comb^i$ |
| Number of flows | $N_F$ |
| Total slots in scheduling length | $N_S$ |
| Schedule length [seconds] | $L_S$ |

For the input flows in Fig. 4, we provide a non-conflicting scheduling sequence as an example using a *benchmark FAT* algorithm here. Let us assume that packets arrive at an Ethernet switch by time 0. *Basic-offset algorithm* assigns the periodic timeslots to the flows 1, 2, and 3: specifically, it assigns to flow 1 timeslots starting at {0, 1.6, 3.2, 4.8} μs, for flow 2 at {0, 3.2} μs, and for flow 3 at {0} μs. However, this solution results in two conflicts, i.e., at times {0, 3.2} μs. These conflicts are resolved by *benchmark FAT algorithm*, by allocating available un-allocated slots to flows involved in each of the conflicts. Flow 1 has schedule of {0, 1.6, 3.2, 4.8} μs, however by the end of the first timeslot, packets from flows 2 and 3 also arrive at the Ethernet switch which results in a conflict. Hence, packets from flow 1 remain unaffected, whereas the other packets get the next unallocated timeslots. The same procedure is applied at timeslot 3.2 μs also, thus resulting in the sequence {0, 1.6, 3.2, 4.8} μs, {0.8, 4.0} μs, and {2.4} μs for the three flows, respectively.

For the same input scenario, C-FIT produces several schedules using different flow orders. For the flow order {1 > 2 > 3}, the schedule is {0, 1.6, 3.2, 4.8} μs for flow 1, {0.8, 4.0} μs for flow 2, and {2.4} μs for flow 3, same as FAT algorithm.

For flow order {1 > 3 > 2}, the schedule is {0, 1.6, 3.2, 4.8} μs for flow 1, {2.4, 5.6} μs for flow 2, and {0.8} μs for flow 3.

For the flow order {2 > 3 > 1}, the schedule is {0, 1.6, 3.2, 4.8} μs for flow 1, {0.8, 4.0} μs for flow 2, and {2.4} μs for flow 3.

For flow order {2 > 1 > 3}, the schedule is {0, 1.6, 3.2, 4.8} μs for flow 1, {0.8, 4.0} μs for flow 2, and {2.4} μs for flow 3.

For flow order {3 > 1 > 2}, the schedule is {0, 1.6, 3.2, 4.8} μs for flow 1, {2.4, 5.6} μs for flow 2, and {0.8} μs for flow 3.

For flow order {3 > 2 > 1}, the schedule is {0, 1.6, 3.2, 4.8} μs for flow 1, {0.8, 4.0} μs for flow 2, and {2.4} μs for flow 3.

All the flow orders in this scenario produced zero jitter, so any schedule can be selected. Since C-FIT approach considers all possible flow orders, the complexity gets exponential. One way to reduce complexity is to stop running the algorithm as soon as a flow order produces zero jitter. Since the fronthaul consists of a limited number of flows multiplexed at the Ethernet switch, the complexity is not a big concern.

## ALGORITHM 1: BASIC-OFFSET ALGORITHM

**Input:** $R_{CoE}^i$, $R_E$, $L_E$ (Assume flows are synchronized at input)
**Output:** Schedule of CoE flows on Ethernet output
**Step 1:** Calculate transmission time for flow i packet on incoming Ethernet link as $T_P^i = \frac{L_E}{R_{CoE}^i}$

Calculate Scheduling length $L_S$ as lowest common multiple of $T_P^i$, i.e., $L_S = lcm\ (T_P^i)$

Calculate outgoing Ethernet timeslot size $T_{ETS} = \frac{L_E}{R_E}$

Calculate number of timeslots in $L_S$ for flow i, $L_{SF}^i = \frac{L_S}{T_P^i}$

**Step 2:** initialize: offset_nf = 0;
  //starting offset value of next flow is set to zero
  **for** i = 1 to $N_F$:
    mat = $comb^i$;
    /* mat is 2-dimensional temporary matrix that holds the contents of $comb^i$ */
    offset = 0;
    **for** j = 1 to $L_{SF}^i$
      in a certain flow
      mat (j, 1) = offset + offset_nf;
        // mat (j, 1) represents start time of packet j
      mat (j, 2) = mat (j, 1) + $T_{ETS}$;

```
                // mat (j, 2) represents end time of packet j
            offset = offset + T_P^j ;
        end
        offset_nf = offset_nf + T_ETS;
        comb^i = mat;
        /* comb^i is 2-dimensional matrix that
                holds start and end time of each packet in flow i*/
    end
```

---

**ALGORITHM 2: COMB FITTING (C-FIT)**

Input:  Schedule of CoE flows given by basic-offset algorithm
Output: Non-conflicting schedule of CoE packets
**Step 1:** Form all possible permutations flow orders
        from 1 to $N_F$ ($N_F!$ different sequences) denoted
        by $SEQ_{PF}^m$, where m = 1 to $N_F!$.
**Step 2: for** each sequence $SEQ_{PF}^m \in \{SEQ_{PF}^m\}$
        **for** j in $SEQ_{PF}^m$
            initialize: matcomb = first element in $\{SEQ_{PF}^m\}$
            matcomb = **matcombine** (comb^j, matcomb)
        **end**
        Calculate jitter matcomb
    **end**
    Pick matcomb with least amount of jitter

**MATCOMBINE SUBROUTINE**

Input: comb^i, comb^j (any two schedules)
Output: Combined non-conflicting schedule
**Step 1:** Initialize: matcomb as a matrix with length as sum lengths
        of comb^i, comb^j
**Step 2:** Take longest sequence out of comb^i, comb^j, and add its
        contents to matcomb, call the other matrix mattemp
**Step 3:** Shift mattemp by multiples of $T_{ETS}$ to form a perfect
        non-conflicting schedule with matcomb
**Step 4: if** success in this procedure
        Copy mattemp to matcomb and **return** matcomb
**Step 5: else**
        Copy non-conflicting packets of mattemp to matcomb
        **for** all conflicting packets in mattemp
            Find nearest open timeslot which can fit the packet
            and update matcomb
        **end**
        **return** matcomb
    **end**

---

**ALGORITHM 3: FIRST AVAILABLE TIMESLOT (FAT)
FOR BENCHMARK**

Input:  Schedule of flows given by basic-offset algorithm
Output: Non-conflicting schedule of CoE packets
**Step 1: for** i = 1 to $N_F$
        initialize: fatcomb = comb^i;
        // fatcomb is temporary matrix that holds contents of
        comb^j
        fatcomb = **fatcombine**(comb^i, fatcomb)
    **end**
    Calculate jitter for fatcomb

**FATCOMBINE SUBROUTINE**

Input: comb^i, comb^j (any two schedules)
Output: Combined non-conflicting schedule
**Step 1:** Initialize: fatcomb as a matrix with length as sum
        lengths of comb^i, comb^j
**Step 2:** Take longest sequence out of comb^i, comb^j, and add its
        contents to fatcomb, call the other matrix mattemp
**Step 3:** Copy non-conflicting packets of mattemp to fatcomb
        **for** all conflicting elements in fattemp
            Find nearest open timeslot which can fit the packet
            and update fatcomb
        **end**
        **return** fatcomb

---

## V. Performance Evaluation and Results

This section presents evaluation of CoE performance metrics such as delay and jitter in the Ethernet fronthaul obtained through Verilog pre-synthesis experiments and simulations.

### A. Ethernet Encapsulation Delay

We present results on the impact of CPRI line rates and payload size on the Ethernet encapsulation delay of CoE flows. We also show how the delay affects the fronthaul distance. An FPGA pre-synthesis verification is performed to map CoE and analyze delay performance of the multi-hop mobile fronthaul [6]. Pseudo random binary sequence (PRBS) data is generated and encapsulated in Ethernet frame using Verilog hardware description language (HDL) and evaluated utilizing ModelSim as a HDL simulator.

Ethernet header consists of 24 bytes, which contains layer-2 Ethernet MAC header fields such as source address (6 bytes), destination address (6 bytes), Ethernet type (2 bytes), RoE header (6 bytes), frame check sequence (4 bytes). The generated data is framed at a clock rate (i.e., a clock cycle takes 6.4 ns) of 10 Gbps. Thus, generating Ethernet header overhead of 24 bytes requires three clock cycles which is 19.2 ns (as shown in Fig. 5). The experiments are conducted for three payload sizes - 500, 1000, and 1500 bytes - to study the effect of payload size on the encapsulation delay.

Figure 5 verifies that Ethernet encapsulation is successfully designed and implemented in Verilog HDL. The left side of the waveform shows the labels of generated data, header fields, Ethernet frame (encapsulating the generated data), and inter-frame gap fields, while the right side shows the corresponding timing information. *data_generation_prbs* shows the generated data in pseudo-random form. *dst_src* in the pre-synthesis evaluation shows the first part of the header containing destination and part of source fields. *src_len_roe_header* field shows the remaining part of the source field, Ethernet type, and first part of roe header. *roe_header_fcs* field shows the remaining part of roe header and frame check sequence (FCS). *roe_payload* field indicates IQ samples encapsulated in Ethernet payload. Note that the data is generated continually at all times including header generation time. The markers show layer-2 Ethernet MAC overhead for a single Ethernet frame as 19.2 ns. For 615 Ethernet frames, the delay would be 19.2×615 ns = 11.8 μs as shown in Table II (for CPRI option 1).

Figure 6 shows the encapsulation delay ($T_{encap}$) as a function of CPRI line rate ($R_{CPRI}$) options with different Ethernet packet payload $P_E$ sizes (500, 1000, and 1500 bytes). As expected, the encapsulation delay decreases as $R_{CPRI}$ increases because, higher $R_{CPRI}$ flow takes shorter time to fill up the Ethernet packet payload size. Moreover, the encapsulation delay decreases as $L_E$ size decreases at the given $R_{CPRI}$ option, as lower payload gets filled in lesser time. Figure 7 shows the distance of fronthaul (based on latency constraints) calculated using Eq. (6), for different CPRI line rates (options 1 to 6) and for the number of Ethernet switches the packets cross for Ethernet payload of 1250 bytes. It shows that as the number of hops increase (more Ethernet switches crossed), the distance of fronthaul decrease due to added delay from the Ethernet switch using store-and-forward mechanism.

The measured values of switch delay are in concurrence with the calculated values (hop delay) as in Table I. Higher line rates also support lower fronthaul distances due to larger number of Ethernet packets that need to be generated for a single 10-ms radio frame. The distances supported using CoE are good for access and metro network coverage distances. Figure 8 shows that a larger Ethernet payload leads to lower CoE overhead, thus supporting longer fronthaul distance.

Hence, CoE can be implemented for different CPRI line rates by compromising a few kilometers in the fronthaul (i.e., from a minimum of 1 kilometer to a maximum of 10 kilometers as a function of the CPRI line rate, and the Ethernet packet payload size as shown in Figs. 7 and 8).

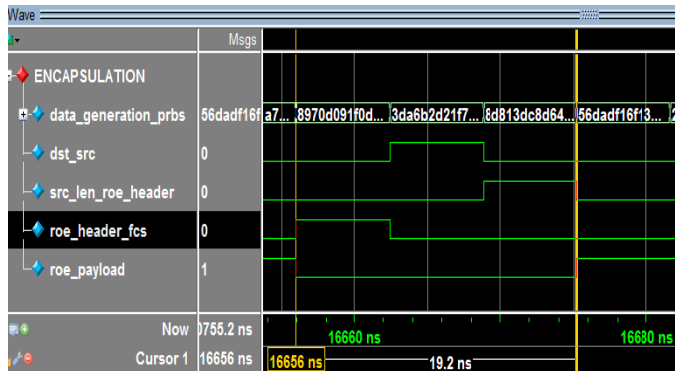

Figure 5: FPGA pre-synthesis simulation of CoE encapsulation.

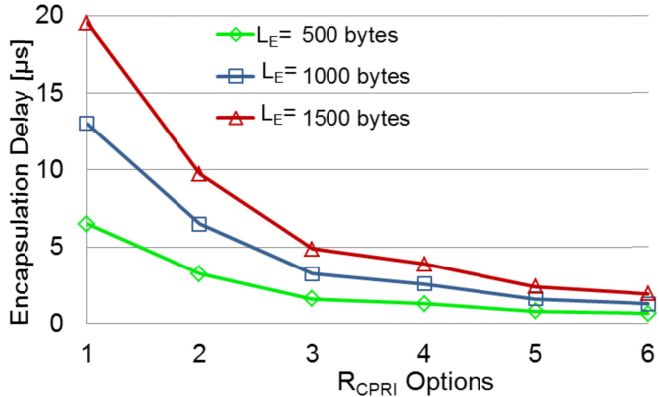

Figure 6: Encapsulation Delay with different $R_{CPRI}$ options and different $L_E$ sizes.

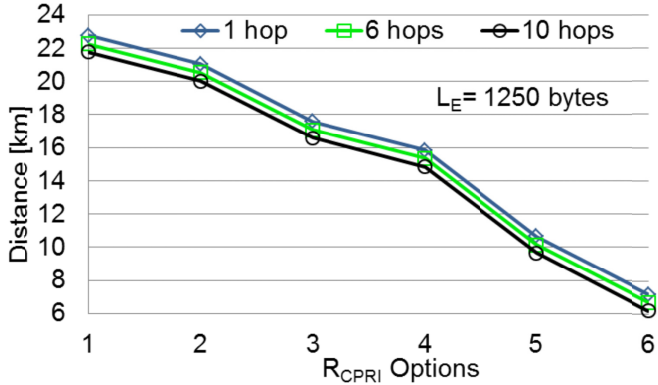

Figure 7: Fronthaul distance supported in multi-hop scenario with $L_E$ =1250 bytes.

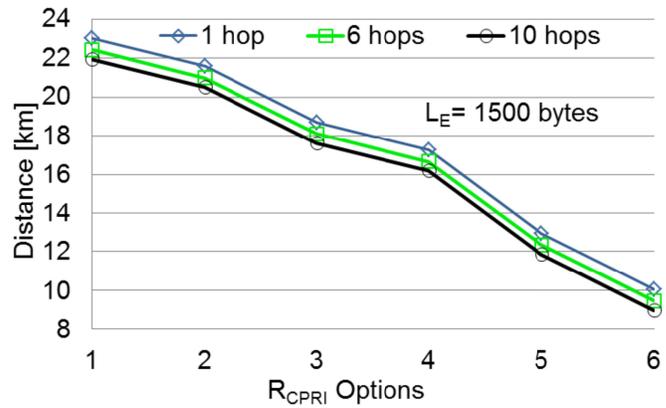

Figure 8: Fronthaul distance supported in multi-hop scenario with $L_E$ =1500 bytes.

### B. Jitter for CoE on Scheduled Ethernet

This section evaluates the performance of C-FIT algorithm, and compares it with the benchmark FAT algorithm. An event-driven simulator built in-house in matlab is used to evaluate both algorithms. The following indexes are considered to evaluate the performance of the algorithms: Load to Ethernet ratio (LER) is defined as the ratio between sum of input CoE-flow rates and the Ethernet rate, and jitter measured at the REC. Over 5000 random combinations of input rates are generated and scheduled over Ethernet and the experiment is repeated 10 times. The value of jitter for particular LER is averaged and plotted. Each plot (Figs. 9-12) simulates random CoE rates derived by encapsulating CPRI flows from line rates uniformly selected from 1 to x (x ≤ 9). Uniformly-distributed random number of flows are multiplexed on an Ethernet link.

Figures 9-10 show the effect of increasing line rates while keeping the number of flows in constant range, whereas Figs. 11-12 show the effect of number of flows multiplexed while keeping the range of line rates constant. Figure 9 shows the value of jitter vs. increasing LER for C-FIT algorithm compared to benchmark FAT algorithm. The CoE rates are uniformly picked from option 1 to option 9, and the number of flows multiplexed are randomly picked from 2 to 5. It can be seen that jitter for C-FIT remains zero until load ratio of 0.3 and then increases. We call the LER until which jitter remains zero and then increases to non-zero value as the inflection point; hence, 0.3 is inflection point in this case. The marked 0.35 LER value shows the maximum allowed jitter of 65 ns. The trend of jitter vs. LER is not only dependent on LER, but also the periodicity of the flows, since more flows that are multiples of each other can form jitter-free schedules without resulting in conflicts. However, if LER is low, there is enough room for the flows to fit in perfectly without conflicts, hence leading to zero jitter.

Figure 10 shows jitter vs. load ratio for CoE rates with range from option 5 to option 9 and number of flows from 2 to 5. We see that the trend is similar till LER of 0.2 (inflection point); then jitter increases. Jitter is below 65 ns until LER of 0.25 for C-FIT. Red dots in the Figs. 9-12 represents 65 ns inflection point. C-FIT shows a monotonically increasing jitter behavior, but benchmark FAT does not. This is because C-FIT explores all possible flow orders and takes the least jitter schedule, but benchmark FAT considers only the given input sequence which can largely affect the jitter values making the relation with LER not as pronounced as expected.

Figures 11-12 show the effect of number of flows multiplexed on the Ethernet switch. Figure 11 has 1-3 flows multiplexed on Ethernet. We see that inflection point is 0.3 and jitter is acceptable until LER of 0.35. Figure 12 shows 4 to 6 flows multiplexed on Ethernet. The inflection point is 0.3 with acceptable jitter until LER of 0.36, and there is monotonic increase in the jitter with higher load ratios. We see from Figs. 11-12 that jitter using FAT algorithm is higher when higher number of flows (i.e., 4 to 6) are multiplexed. This is because, as fronthaul topology gets larger (more number of flows), the number of conflicts increases.

Although results indicate that only 30% of the average amount of Ethernet bandwidth can be used if we want to satisfy jitter requirement to multiplex CoE flows, there are certain flow combinations which can achieve very low jitter with much higher Ethernet utilization. In fact, input flows that are multiples of each other are found to achieve higher Ethernet utilization while guaranteeing tolerable jitter of 65 ns. Moreover, it is worth noting that the redundant Ethernet capacity could be utilized to send other non-time-sensitive data, as in [28] where fronthaul, midhaul, backhaul can share Ethernet capacity.

The encapsulation delay of CPRI packets decreases as line rate increases as shown in Table II; however, jitter increases as LER increases. Hence, there is a need to make a careful choice of Ethernet rate for a given CoE topology (rates).

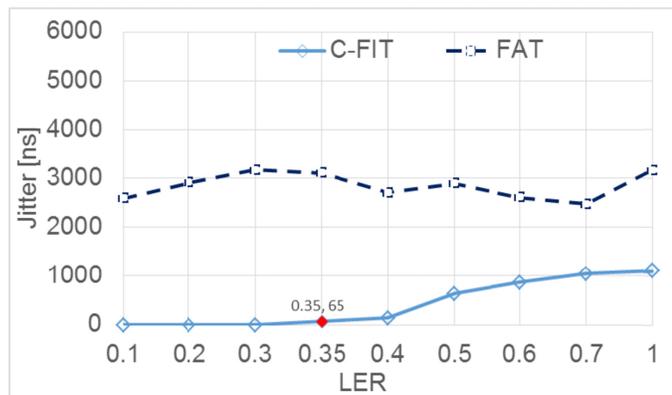

Figure 9: Jitter vs. Load to Ethernet Ratio for number of flows 2 to 5 and line rates 1 to 9.

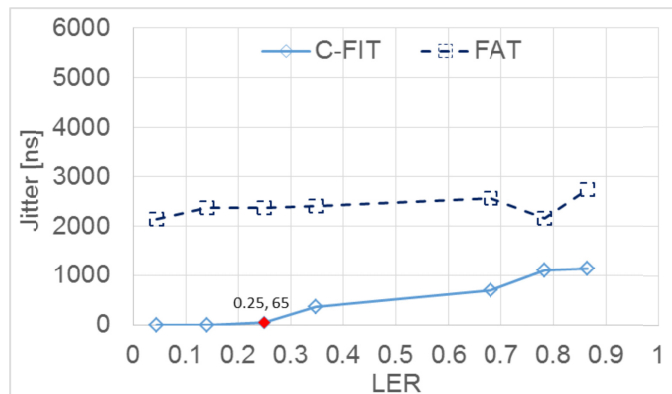

Figure 10: Jitter vs. Load to Ethernet Ratio for number of flows 2 to 5 and line rates 5 to 9.

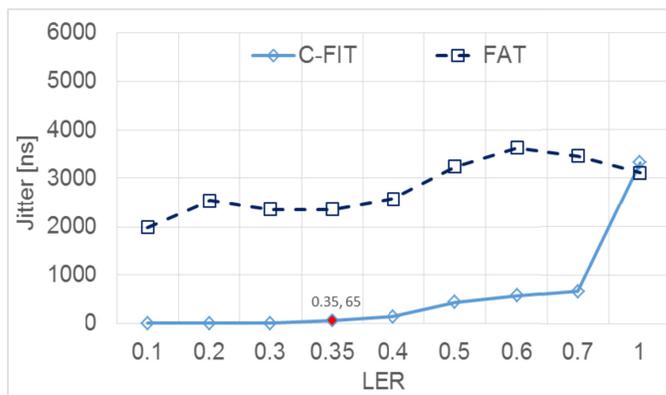

Figure 11: Jitter vs. Load to Ethernet Ratio for number of flows 1 to 3 and line rates 1 to 9.

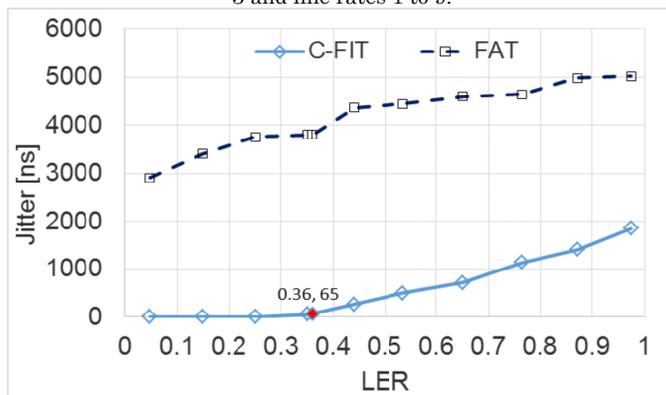

Figure 12: Jitter vs. Load to Ethernet Ratio for number of flows 4 to 6 and line rates 1 to 9.

## VI. CONCLUSION

Ethernet fronthaul is expected to provide many benefits to mobile networks such as 5G. CPRI over Ethernet (CoE) can be a cost-efficient alternative to CPRI fronthaul, as Ethernet is easily available, and can be a stepping stone to many useful applications. In this study, we showed that CoE encapsulation and switching introduces a slight delay which can compromise a few kilometers in the multi-hop mobile fronthaul. Moreover, jitter was studied in terms of load to Ethernet ratio (LER), number of flows, and flow combination. For a given topology, a scheduling method that completely eliminates jitter can be provided by using certain Ethernet rate with the proposed comb-fitting (C-FIT) scheduling. Proposed C-FIT scheduling performs considerably well compared to benchmark first-available-timeslot (FAT) algorithm. In particular jitter reduction as big as units of microseconds can be achieved.